\documentclass[11pt]{article}
\usepackage{graphics}

\begin{document}

\begingroup
\thispagestyle{empty}
\baselineskip=14pt
\parskip 0pt plus 5pt
\begin{center}
{\large EUROPEAN ORGANIZATION FOR NUCLEAR RESEARCH}
\end{center}

\begin{flushright}
CERN--EP/2000--117\\
DSF--26--2000\\
August 4, 2000 \\
\end{flushright}
\bigskip

\begin{center}
\Large
\textbf{A search for $Z'$ in muon neutrino associated charm production}
\end{center}

\bigskip\bigskip
\begin{center}
{ P.~Migliozzi${^a}$\footnote{On leave of absence from INFN, 
Sezione di Napoli, Italy.}, G.~D'Ambrosio$^b$, 
G.~De Lellis${^b}$\footnote{Partially supported by the 
``Fondo Sociale Europeo''.}, 
F.~Di Capua$^b$, G.~Miele$^b$, P.~Santorelli$^b$}\\
\end{center}

\vskip 1.0cm
\begin{center}
\small
a) CERN, Geneva, Switzerland.
\vskip 0.1cm
b) Universit\`a ``Federico II'' and INFN, Napoli, Italy.
\end{center}
\bigskip\bigskip\bigskip

\begingroup
\leftskip=0.4cm
\rightskip=0.4cm
\parindent=0.pt
\begin{center}
\begin{abstract}
In many extensions of the Standard Model the presence of an extra neutral 
boson, $Z^{\prime}$, is invoked. A precision study of weak neutral-current 
exchange processes involving only second  generation fermions is still 
missing. We propose a search for $Z^{\prime}$ in muon neutrino associated 
charm production. This process only involves $Z^{\prime}$ couplings with 
fermions from the second generation. An experimental method is thoroughly 
described using an {\it ideal detector}. As an application, the  
accuracy reachable with present and future experiments has been estimated. 
\end{abstract}
\end{center}
\endgroup
\vskip 0.4cm
\vfill
\begin{center}
(Submitted to {\it  Physics Letters B})
\end{center}

\endgroup

\newpage

\section{Physics motivation}
Since its experimental confirmation, the Standard Model of electroweak 
interactions (SM) has been challenged in all possible 
directions  (see Ref.~\cite{PDG00} and references therein) 
with the aim of finding the signature of new fundamental physics. 
Although at  present there is no clear evidence of any departure from  it, 
sometimes unexpected deviations show up in experiments, for instance 
the anomalies seen at LEP in $R_{b}$ \cite{PDG96}, at HERA 
at high  $Q^{2}$ \cite {H1-Zeus} and at Tevatron \cite{CDF}. 
Consequently ad hoc models are built.
However, none of these discrepancies has survived to further 
experimental investigations. Recently, in Atomic Parity Violating 
experiments, a discrepancy from the SM prediction 
has been observed \cite{APV}. This could be explained in terms
of extra $Z$ bosons \cite{APVth}. 

The existence of  $Z'$ boson is foreseen in many extensions of 
the SM and is associated with extra $U(1)$ gauge symmetries. For
instance, 
in the symmetry breaking pattern of $E_6$ or $SO(10)$ the $Z'$ boson is 
contained in the low energy extension of the SM-like $SU(2)_R \times 
SU(2)_L \times U(1)_{B-L}$ or $SU(2)_{L} \times U(1) \times U(1)$, see 
Ref.~\cite{gut} and references therein.

ALEPH and OPAL experiments have put limits on the presence of
$Z^{\prime }$ by studying the contribution of new contact interactions
in the processes $e^+ e^- \rightarrow f \overline{f}$ \cite{LEP}.
The CHARM II experiment derived constraints on additional $Z$ bosons 
from $\nu_\mu e \rightarrow \nu_\mu e$ scattering measurements
\cite{charmii}.
These searches assume a family independent scheme for the $Z'$ couplings
to leptons and quarks. Moreover, there exist severe constraints 
in the first two generations 
on FCNC $Z^{\prime}$ from  $K_L - K_S$ mass splitting and on
lepton family violating  $Z^{\prime}$ from $B(\mu \rightarrow 3 e)$. A
diagonal $Z^{\prime}$  strongly coupled to the second family could be 
limited by  $J/ \psi\rightarrow \mu^+ \mu^-$. However,  the pure
electro-magnetic contribution and the hadronic uncertainties weaken this limit.
Constraints on
a $Z^{\prime}$ which couples differently only to the third generation are
somewhat 
weaker \cite{PDG00}.

The large mass difference between the top quark and the remaining ones has 
recently suggested a new class of models based on $SU(N) \times SU(N)$. 
In this framework the large mass difference can be naturally accommodated 
as well as the well-known phenomenology of weak interactions. 
Moreover, due to the extended gauge interaction the $Z'$ presence
in all processes involving the third family could be enhanced
\cite{third}. 

A precision study of weak neutral-current exchange processes 
involving only  second generation fermions is still missing. Therefore, 
it is mandatory to test the SM predictions in this sector. 

In this letter we propose a search for $Z'$ 
through the measurement of associated charm production induced by
$\nu_\mu$  neutral-current interactions $(\nu_\mu + N \rightarrow \nu_\mu 
+ X + c \bar{c})$. An ideal detector is exploited to perform this
measurement. 
The importance of such a search is twofold since on one hand this 
performs a further test of SM family universality and on the other hand 
one can check the presence of possible $Z'$ mainly coupled to the second 
and/or third families. It is worthwhile stressing that the proposed search 
is model independent. 

We shall show that  present neutrino experiments already 
constrain extensions of the Standard Model. Nevertheless, since
neutrino factories could become a real perspective for the future, 
it is conceivable that the new generation of $\nu $--experiments will 
be able to probe new physics with higher sensitivity. 

\section{Four fermions contact terms and extra neutral bosons}
\label{sec:contact}
At $Q^2 \ll M_Z^2$ 
the neutral-current effective Lagrangian ruling the associated charm 
production induced by $\nu_\mu$  is given by
(see for example Ref.~\cite{PDG00} and references therein for notation)
\begin{equation}
{\cal L}^{\nu c \overline{c}}_W = - \frac{G_F}{\sqrt{2}} ~
\overline{\nu}_\mu \gamma ^{\alpha }(1-\gamma _{5})\nu_\mu
\left[ \epsilon_L(u) ~ \overline{c}\gamma_\alpha (1 - \gamma_5)c +
\epsilon_R(u) ~\overline{c} \gamma_\alpha (1 + \gamma_5) c
\right],
\label{lw}
\end{equation}
where the parameters $\epsilon_L(u)$ and $\epsilon_R(u)$ account for the 
different coupling of left-handed and right-handed {\it up}-kind quarks 
to neutral-current respectively. Theoretically these two parameters are
very precisely predicted \cite{PDG00}, namely
\begin{equation}
\epsilon_L^{th}(u)= 0.3459\pm 0.0002 ~~~, ~~~~~~~~~~
\epsilon_R^{th}(u)=-0.1550 \pm 0.0001~~~.
\label{epsth}
\end{equation}
Since in the SM the matter-gauge coupling is family independent,
the experimental determinations of the above parameters (\ref{epsth})
are obtained by looking at processes where the four fermions
involved 
come from first generation only or from two different families, as 
for instance in the ratio $R_q= \sigma(e^- e^+ \rightarrow q
\overline{q})/
\sigma(e^- e^+ \rightarrow \mu^- \mu^+)$. This experimental knowledge, 
which has not yet reached the accuracy of the theoretical predictions,
gives \cite{PDG00}
\begin{equation}
\epsilon_L^{ex}(u)= 0.330\pm 0.016 ~~~, ~~~~~~~~~~
\epsilon_R^{ex}(u)=-0.176^{+0.014}_{-0.006}~~~,
\label{epsex}
\end{equation}
which are in 1 $\sigma$ agreement with theoretical values. Nevertheless
pure measurements of these parameters, with a comparable level of
precision, 
in processes involving only the second family are still missing.

The $Z'$ boson presence in the 
process $\nu_\mu + N \rightarrow \nu_\mu + X + c \bar{c}$ can be
introduced 
in a model independent way by the effect of four fermion contact
interactions with new couplings \cite{peskin}.

Muon neutrinos   produced in weak meson decays are left-handed. Therefore,  
the most general SM-like term describing the additional interaction is

\begin{equation}
{\cal L}_{NP}^{\nu c \overline{c}} = 
- \frac{G_F}{\sqrt{2}} \left(M_Z^2 \over M_{Z'}^2\right)
\overline{\nu}_\mu \gamma ^{\alpha }(1-\gamma _{5})\nu_\mu
\left[ \eta_L~ \overline{c}\gamma_\alpha (1 - \gamma_5)c +
\eta_R ~\overline{c} \gamma_\alpha (1 + \gamma_5) c
\right].
\label{np}
\end{equation}
where  $NP$ stands for New Physics, $\eta_R$ and $\eta_L$ are 
the $\nu_{\mu}-c$ new couplings and $M_{Z\prime}$ is the extra boson mass.

Given the additional contribution, the total effective Lagrangian,
${\cal L}_{T}^{\nu c \overline{c}}$, takes the form
\begin{equation}
{\cal L}_{T}^{\nu c \overline{c}} = 
- \frac{G_F}{\sqrt{2}}
\overline{\nu}_\mu \gamma ^{\alpha }(1-\gamma _{5})\nu_\mu~
\overline{c}\gamma_\alpha \left[\epsilon_V(c) - \epsilon_A(c) \gamma_5
\right]c~~~,
\label{eq:T}
\end{equation}
where
\begin{eqnarray}
\epsilon_V(c)&=& \epsilon_L(u)+\epsilon_R(u)   
+ \left(M_Z^2 \over M_{Z'}^2\right) (\eta_L + \eta_R ) \nonumber\\
&\equiv &  \epsilon_V(u)+ \left(M_Z^2 \over M_{Z'}^2\right) 
\eta_V = \left[ 1+ \left(M_Z^2  \over M_{Z'}^2\right) x \right]
\epsilon_V(u)~~~,
\label{epsv}\\
\epsilon_A(c)&=& \epsilon_L(u)-\epsilon_R(u) 
+ \left(M_Z^2 \over M_{Z'}^2\right)
(\eta_L - \eta_R ) \nonumber\\
&\equiv&  \epsilon_A (u) + \left(M_Z^2 
\over M_{Z'}^2\right) \eta_A = \left[1+ \left(M_Z^2  \over M_{Z'}^2\right) 
y\right]
\epsilon_A(u)~~~,
\label{epsa}
\end{eqnarray}
and the parameters $x$ and $y$ give the departure from SM predictions.

\section{Present available data on $\nu_\mu$ associated charm
production}
\label{sec:e531}
The available data on neutrino associated charm production are scarce.
Only one event consistent with the neutral-current production of a pair of
charmed particles has been observed by the E531 Collaboration in an
emulsion hybrid experiment \cite{e531}. 
This event allowed the determination of the associated charm production rate 
with respect to neutral-current production:

\begin{equation}
\frac{\sigma(\nu_{\mu} N \rightarrow c\bar{c} \nu_{\mu} X)}
{\sigma(\nu_{\mu} N \rightarrow \nu_{\mu} X)} =  0.13^{+0.31}_{-0.11}\%.
\end{equation}
No event has been found in charged-current production. Under the
assumption that the primary muon was not identified, the previous result
can be translated into an upper limit at $90\%$C.L. on associated charm
production in the charged-current production of

\begin{equation}
\frac{\sigma(\nu_{\mu} N \rightarrow c \bar{c} \mu X)}
{\sigma(\nu_{\mu} N \rightarrow \mu  X)} \leq 0.12 \%.
\end{equation}

\section{Simulation of the process}
\label{sec:simul}

\begin{figure}[htbp]
\begin{center}
\resizebox{.8\textwidth}{!}{\includegraphics
{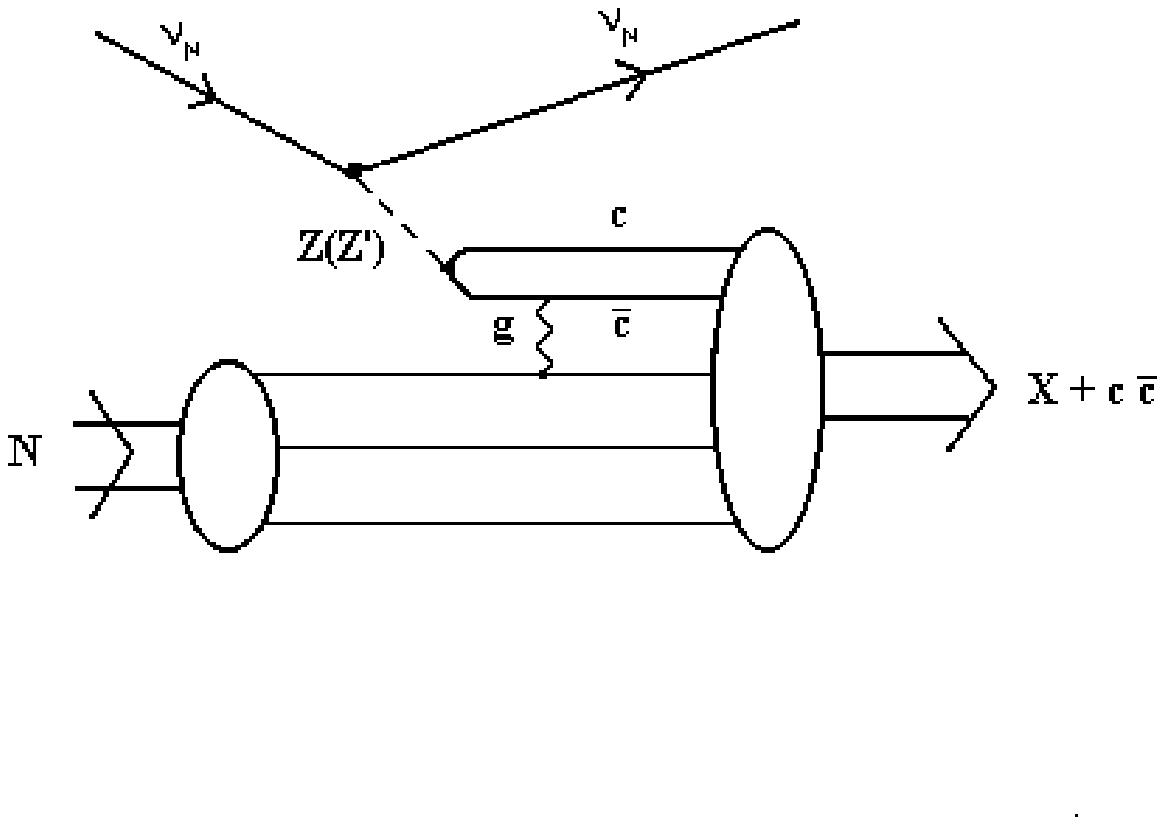}}
      \caption{The boson gluon fusion process diagram 
in $\nu_{\mu}$ interactions.}
      \label{fig:diag}
\end{center}
\end{figure}

The Lagrangian ${\cal L}_{T}^{\nu c \bar{c}}$ defined in equation 
(\ref{eq:T})
contributes to the process $\nu_\mu + N \rightarrow \nu_\mu + X + c
\bar{c}$ 
where charm quarks adronize through the gluon exchange with the nucleon 
partons (boson gluon fusion), see figure \ref{fig:diag}. 

At the relevant $Q^2$ values ($\leq 20 (GeV/c)^2$), the deep inelastic 
scattering phenomenology is very well described by the three flavour
scheme ($u$, $d$, and $s$), see for example Ref.~\cite{tung}.
This implies that the 
sea charm-parton component is negligible in this $Q^2$ range. Therefore
the only process producing a $c\bar{c}$ pair in the final state is the 
boson gluon fusion. 

In order to simulate the process, we have used the HERWIG event generator 
\cite{herwig}. It is based on perturbative QCD calculations and 
provides a good description of all available data at LEP and Tevatron 
\cite{qcd}. 
All final state 
particles are generated and the cross-section value is also computed. 
An associated charm production rate with respect to the neutral-current 
production of $(0.403\pm0.004)\%$ \footnote{The error is only statistical.} 
is predicted by HERWIG. It is consistent 
with the experimental measurement given in Section \ref{sec:e531}. 

\section{Description of the method }
\label{method} 

The search presented in this letter exploits the peculiar topology of the 
associated charm production in $\nu_{\mu}$ neutral-current interactions: 
two charmed hadrons in the final state. Consequently, there are no 
other physical processes which may mimic it. 

Experimentally we are sensitive to the ratio 
\begin{equation}
R = \frac{\sigma_{c \bar{c}}^{NC}}{\sigma^{CC}}
\end{equation}
which can be written as the product 

\begin{equation}
\label{eq:rapporto}
R = \frac{\sigma_{c \bar{c}}^{NC}(Z^{0} + Z^{\prime})}
{\sigma_{c \bar{c}}^{NC}(Z^{0})} 
\times \frac{\sigma_{c \bar{c}}^{NC}(Z^{0})}{\sigma^{CC}} = r \times f
\end{equation}
where $\sigma_{c \bar{c}}^{NC}(Z^{0})$ is the cross-section of the
associated 
charm production process in $\nu_{\mu}$ interactions in absence of the 
$Z^{\prime}$ boson, $\sigma_{c \bar{c}}^{NC}(Z^{0}+ Z^{\prime})$ includes
the 
contribution of the new neutral boson and $\sigma^{CC}$ is the $\nu_{\mu}$ 
deep inelastic charged-current cross-section.

In the following we assume a 50 GeV mono-energetic 
$\nu_{\mu}$ beam 
\footnote{
The results achievable with a real neutrino spectrum of mean 
energy $<\!\!E_{\nu}\!\!>$ are  rather well reproduced  by 
using a simple mono-energetic beam 
with energy equal to $<\!\!E_{\nu}\!\!>$. 
}. 
In the following we assume a 50 GeV mono-energetic $\nu_{\mu}$ beam. 
Under this assumption by using the simulation program described in Section 
\ref{sec:simul} the ratio $f$ results to be $(1.25 \pm 0.01) \times
10^{-4}$. 
From equation (\ref{eq:rapporto}) it is then clear that the only relevant 
contribution is coming from the ratio $r$. 
 
If we parameterise the ratio $r$ in terms of the $x$, $y$ and
$M_{Z\prime}^2$ variables defined in Section \ref{sec:contact},  
the most general expression we get is:

\begin{equation}
r(x,y,M^2_{Z^{\prime}}) = 
1 +  \left(500  \over M_{Z'}^2\right)^2 ( A_1 y + B_1 x) 
 +  \left(500  \over M_{Z'}^2\right)^4 
(A_2 y^2 +  B_2 x^2 + C_1 xy) .
\label{eq:fitfor}
\end{equation}
Fitting the data from the simulation with the previous function,  
the values of the coefficients we get are: $ A_1 = 0.1, A_2 = 0.003, 
B_1 = 0.02, B_2 = 0.0007 $ and $C_1 = -0.0002$. 
The fit is valid in the $[-30, 30]$ range for both $x$ and $y$ variables.

\begin{figure}[htbp]
\begin{center}
\resizebox{.6\textwidth}{!}{\includegraphics
{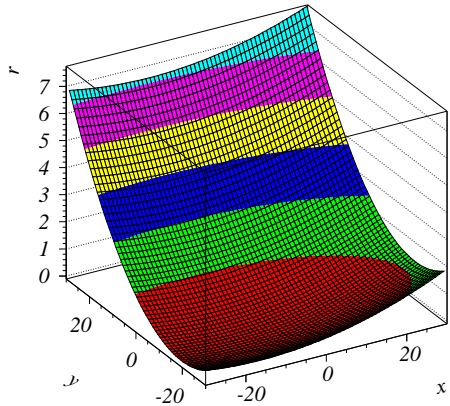}}
      \caption{The ratio $r$ is plotted by assuming 50 GeV $\nu_{\mu}$ 
energy and $M_{Z^{\prime}}$ = 500 GeV$/c^2$. }
      \label{fig:r50b}
\end{center}
\end{figure}

In Figure~\ref{fig:r50b}  the fitted function $r$  for $M_{Z^{\prime}} = 500$ 
GeV$/c^2$ is shown. 

The number of observed events, $N_{S}$, can be written as

\begin{equation}
N_{S} = N_{c \bar{c}} \cdot \frac{\varepsilon_S}{\varepsilon_B} \cdot  r
\label{eq:events}
\end{equation}
where $N_{c \bar{c}}$ is the number of observed events without the 
$Z^{\prime}$ effect, $\varepsilon_S$ and $\varepsilon_B$ are 
the reconstruction efficiencies for the events with and
without a $Z^{\prime}$, respectively. 

\subsection{Measurement accuracy in an ideal detector}
We assume an ideal detector designed to identify charmed mesons and 
barions which travel on average about 1 mm before decaying 
if produced by 50 GeV neutrinos. 
In order to obtain this goal we need a very high 3D resolution 
tracker. Nuclear emulsions  have the required spatial resolution (less
then 1 $\mu$m). A good hadron spectrometer for additional kinematical
analysis 
and a calorimeter to measure the hadronic shower produced in the
interaction are also needed. 
A muon spectrometer in the down-stream part of the apparatus 
will allow us to tag charged and neutral-current interactions. 
It could also be useful to analyse the exclusive semi-leptonic decay
channel.

Once  the charmed particles have been tagged, the $Z^{\prime}$ effect
would show up as an excess/defect of double charmed events in neutral-current
interactions.
If no excess/defect is found it will turn into a limit on the coupling 
parameters. 

The detection efficiencies have been calculated with the cuts defined in 
Table \ref{tab:effcut}. In particular we assume to detect tracks with  
angles  less than 400 mrad. Moreover, in the single prong decays we 
require the minimum 
kink angle to be 15 mrad. A minimum flight length cut of 10 $\mu$m is also 
assumed to distinguish between primary and secondary vertices. 

\begin{table}[htbp]
\small
\begin{center}
\caption{\small Reconstruction efficiencies in the emulsion target. Notice
that the kink angle cut is only applied for single prong decays.} 

\vspace{5mm}
\begin{tabular}{||c|c|c||} \hline
\hline
Cuts &  $\varepsilon_S$ (\%) & $\varepsilon_B$ (\%)  \\
\hline
\hline
Angular cut ($\vartheta$  $\leq$ 0.4) & $87.3\pm0.3$ & $87.5\pm0.3$  \\ 
\hline
Kink angle cut ($\geq$ 15 mrad ) 
& $95.2\pm0.2$   
 &   $95.0\pm0.2$ \\ 
\hline
Flight length cut ($\geq$ 10 $\mu$m) & $95.2\pm0.2$ & $95.4\pm0.2$ \\ 
\hline
\end{tabular}
\label{tab:effcut}
\end{center}
\end{table}

The topology of the two samples of events is extremely similar so that the 
detection efficiencies are the same within the error, as shown in Table 
\ref{tab:effcut}.

In Table \ref{tab:branch} we report the hadronization fractions as predicted 
by the event generator model. No dependency of the hadronization fractions 
on the $Z^{\prime}$ couplings has been observed in this model.

\begin{table}[htbp]
\small
\begin{center}
\caption{\small The hadronization fractions are reported for associated charm 
production events induced by 50 GeV $\nu_{\mu}$. The error is only 
statistical.}

\vspace{5mm}
\begin{tabular}{||c|c|c|c|c||} \hline
\hline
$f$(\%) &  $D^+$ & $D^0$ & $D^{+}_s$ & $\Lambda^{+}_c$  \\
\hline
$D^-$ & $5.3\pm0.2$  & $7.5\pm0.3$  & $0.09\pm0.03$  & $14.8\pm0.4$  \\
\hline
$\bar{D}^0$ & $11.6\pm0.3$  & $15.5\pm0.4$  & $0.32\pm0.06$  & $27.2\pm0.5$ \\
\hline
$D^{-}_s$ & $2.2\pm0.2$  & $2.7\pm0.2$  & $0.10\pm0.03$  &  $6.2\pm0.2$ \\
\hline
$\bar{\Lambda}^{-}_c$ & $1.2\pm0.1$  & $2.3\pm0.2$  & $0.04\pm0.02$  & 
$2.8\pm0.2$ \\
\hline
\end{tabular}
\label{tab:branch}
\end{center}
\end{table}

In Figure~\ref{fig:r50b} we see that for ``large'' $Z^{\prime}$ couplings, 
i.e.~$x$ $\mbox{and}$ $y> 20$, we can get an enhancement of the associated
charm 
production of about a factor  seven. 

\begin{figure}[hbtp]
\vspace{0.5cm}
  \begin{center}
    \rotatebox{0.0}{
      \resizebox{1.0\textwidth}{!}{
\includegraphics{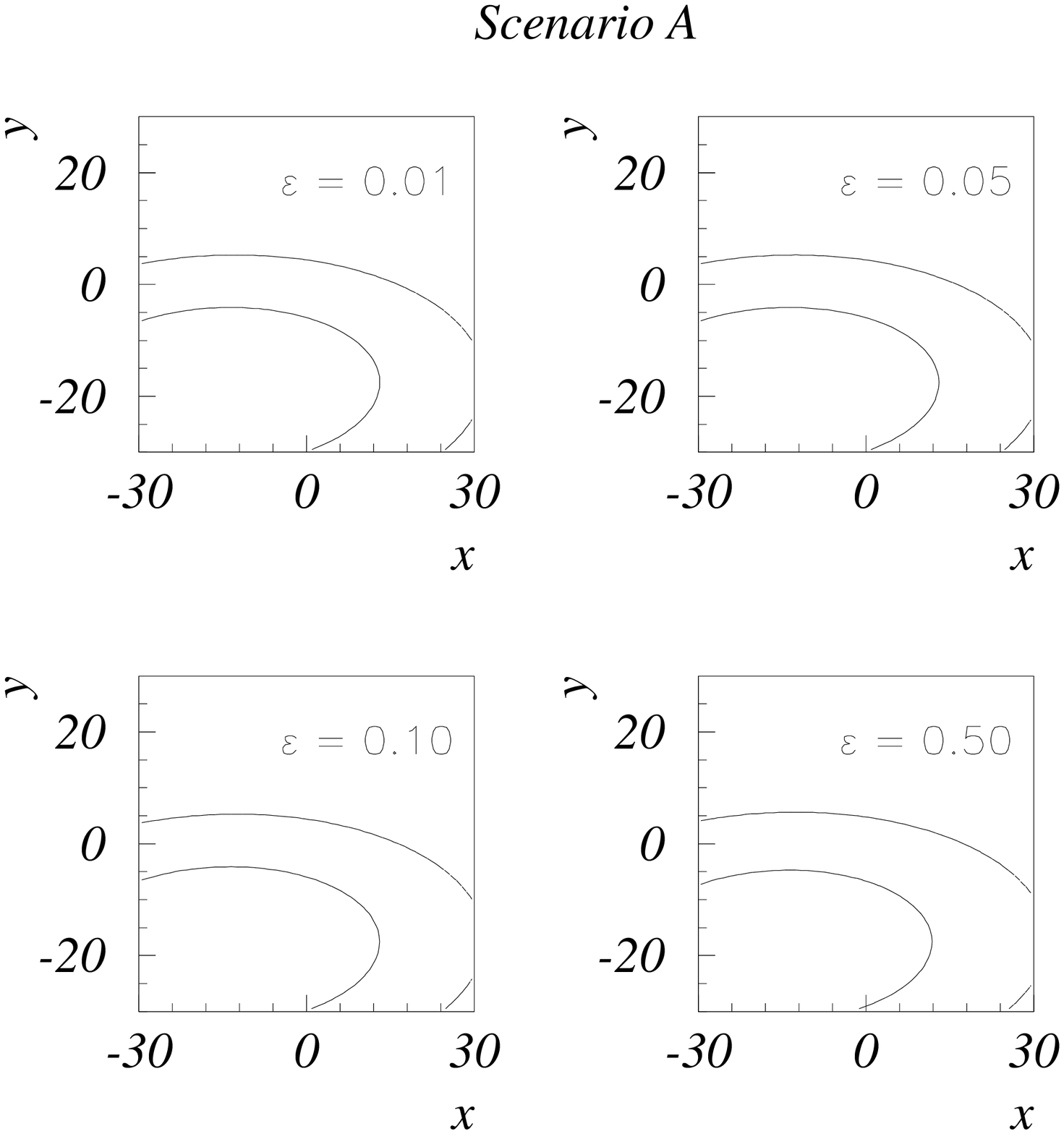} 
\includegraphics{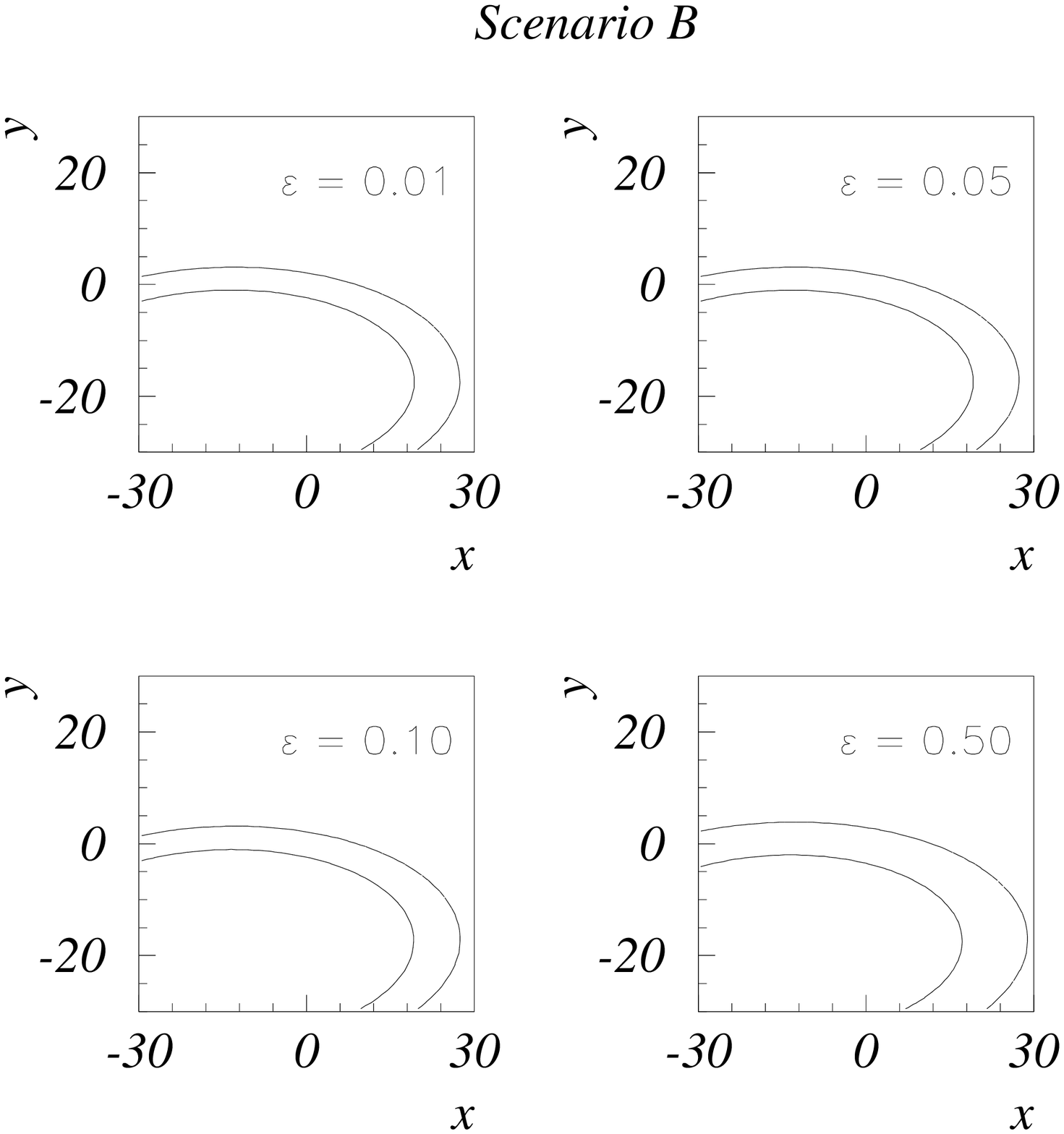}}}
    \rotatebox{0.0}{
      \resizebox{1.0\textwidth}{!}{
        \includegraphics{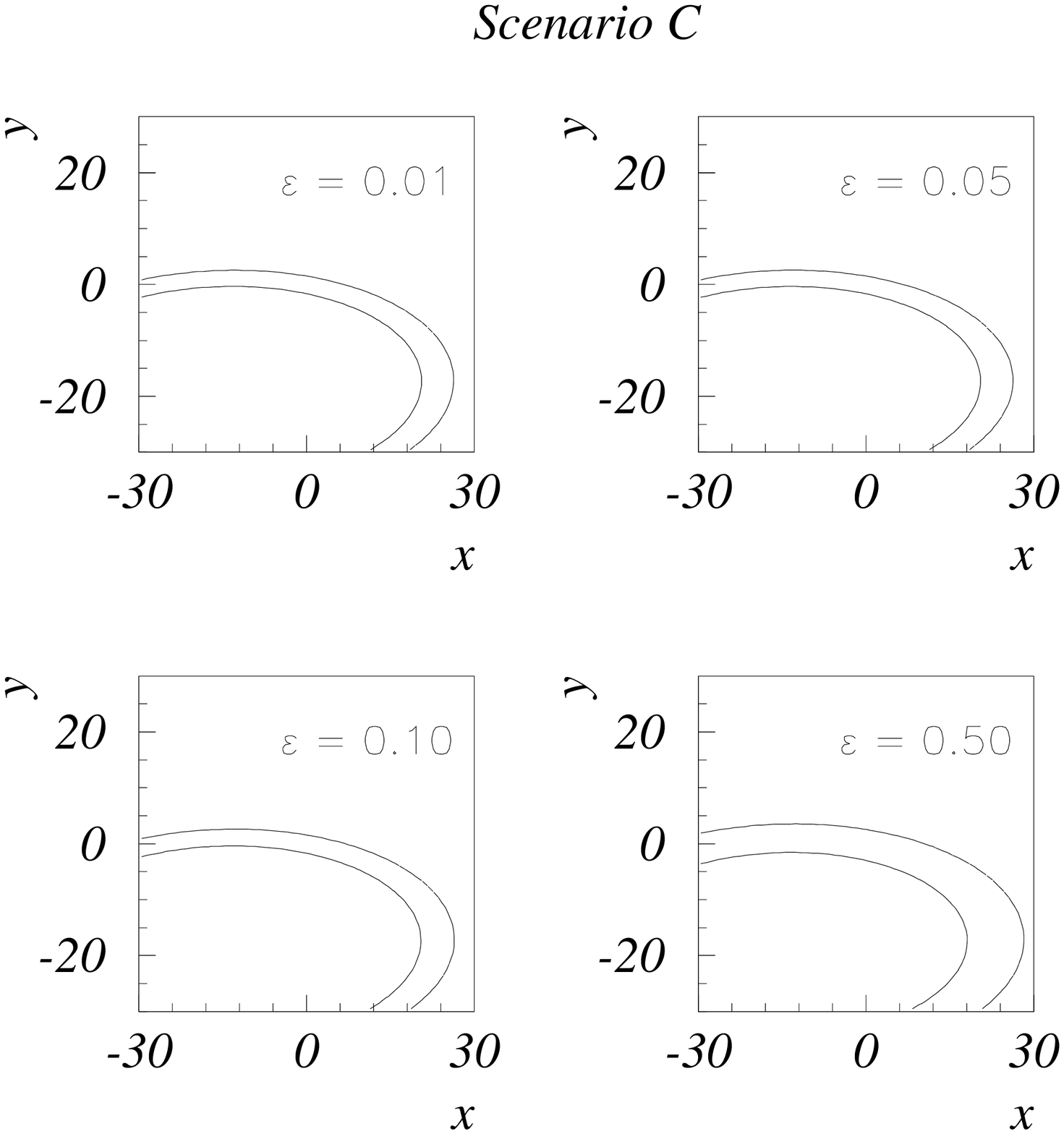}
        \includegraphics{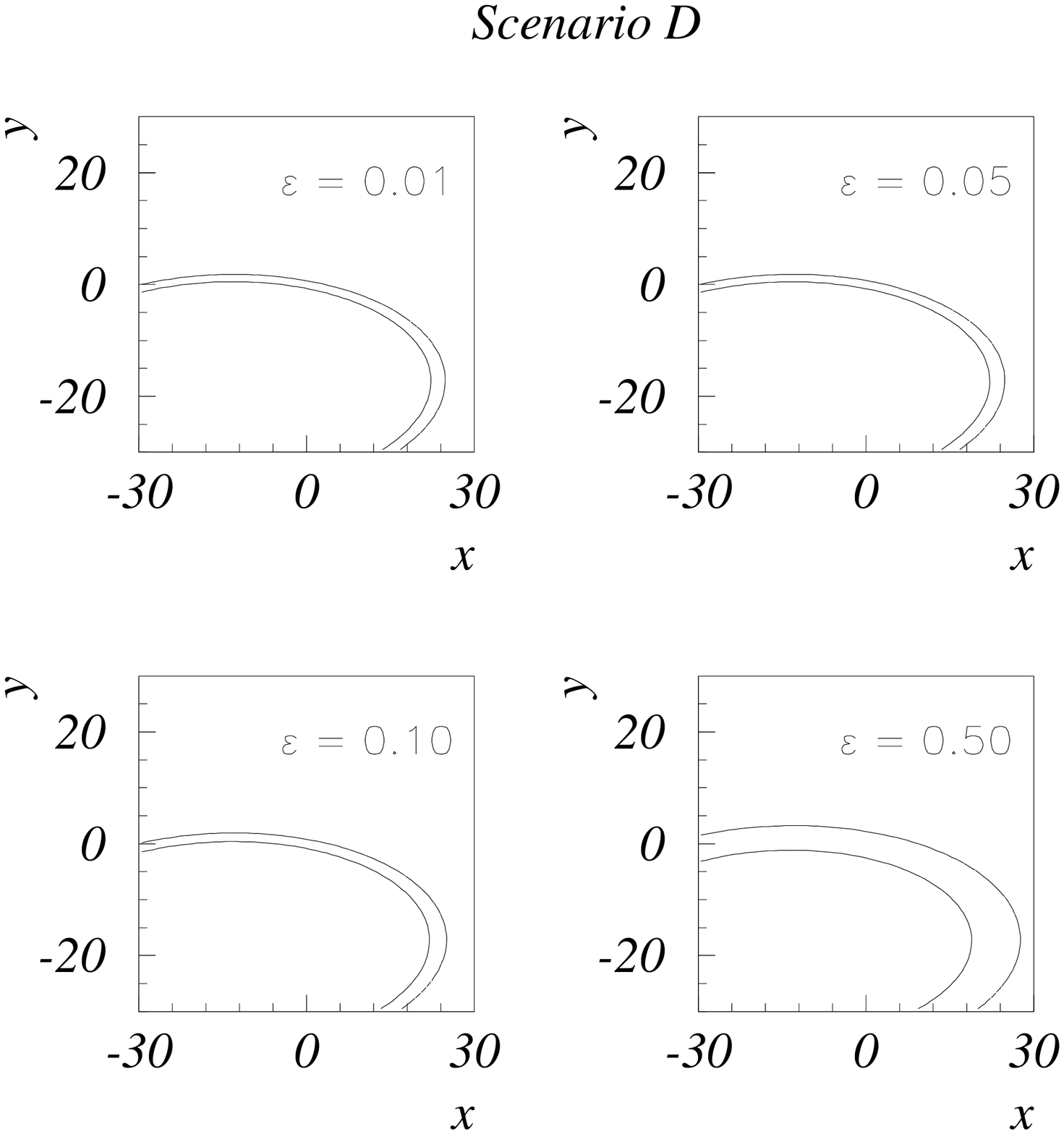}}}
        \caption{\small The sensitivity plots for the $x$ and $y$ variables at 
$M_{Z^{\prime}}$ = 500 GeV$/c^2$ are shown in the four different scenarios 
described in the text.  $\varepsilon$ indicates the systematic error.}
   \label{fig:cont50ev}
  \end{center}
\end{figure}

On the other hand, if we do not observe any excess/defect we can put a
limit on the $x$ and $y$ parameters. As an example we report in 
Figure~\ref{fig:cont50ev} the sensitivity plot at 90\% C.L.
for the $x$ and $y$ variables  at $M_{Z^{\prime}}$ = 500 GeV$/c^2$. 
Different statistics of associated charm production events as well as 
different systematic errors are assumed. In Table \ref{tab:summ} we 
report the summary of the four different scenarios considered 
in Figure \ref{fig:cont50ev}. Each scenario corresponds to a given number of 
associated charm events, $N_{c \bar{c}}$, namely 10, 50, 100 and 500. 
For the sake of simplicity we also report the corresponding number of 
charged current neutrino interactions, $N_{\mu}$. For each scenario the 
systematic error has been ranged from 1\% to 50\%.

The allowed region of parameters is obtained from the formula 

\begin{equation}
1-1.64\cdot \frac{\sigma}{N_{c\bar{c}}} \leq 
\frac{\varepsilon_S}{\varepsilon_B} \cdot  r \leq 1+1.64 \cdot 
\frac{\sigma}{N_{c\bar{c}}} 
\end{equation}
where $\sigma$ is defined as 
\begin{equation}
\sigma = (\varepsilon^{2}_{stat}+\varepsilon^{2}_{sys})
\end{equation}
and includes the error on the event counting from both a statistical and 
systematics source. The factor 1.64 takes into account the required 
confidence level. Therefore in Figure~\ref{fig:cont50ev} for each plot 
the two lines bound the region of coupling parameters where no significant 
excess/defect of associated charm production events is found. In other words, 
an observation of a number of charm pair events in 
agreement with SM predictions excludes the regions outside the band. 

As expected, in the Scenario $A$ the statistical fluctuation is dominant with 
respect to the systematic error so that the bounds are rather large and 
systematics-independent.  On the contrary, for a large statistic
experiment as predicted in Scenario $D$ 
the systematic uncertainties would play a crucial role: the smaller the 
systematic error is, the narrower the allowed parameters band becomes.  

\begin{table}[htbp]
\small
\begin{center}
\caption{\small Summary of the statistics used for the different scenarios
shown in Figure~\ref{fig:cont50ev}. The set of systematical errors used
is also shown.} 
\vspace{5mm}
\begin{tabular}{||c|c|c|c||} \hline
\hline
Scenario & $N_\mu$ & $N_{c\bar{c}}$ & $\varepsilon_{syst}(\%)$  \\
\hline
\hline
$A$ & $1\times10^5$ & $10$ & $1,5,10,50$ \\
\hline
$B$ & $5\times10^5$ & $50$ & $1,5,10,50$ \\
\hline
$C$ & $1\times10^6$ & $100$ & $1,5,10,50$ \\
\hline
$D$ & $5\times10^6$ & $500$ & $1,5,10,50$ \\
\hline
\end{tabular}
\label{tab:summ}
\end{center}
\end{table}

\section{Measurement accuracy with present and future experiments 
statistics}

Among the neutrino experiments which are currently taking or analysing
data, CHORUS\cite{chorus}, which uses nuclear emulsions as a target, has an
adequate spatial resolution to search for associated charm production induced
by muon neutrinos. Starting from a sample of approximately 500000
charged-current events, it is estimated that $\sim350000$ events will be
analysed in the emulsion\cite{chorus2}. Assuming a $50\%$ efficiency to detect
the charmed pair, a statistic of about $20$ events can be
expected. Consequently, the CHORUS experiment can explore the $x$ and $y$
parameter region similar to the one shown in the Scenario $A$ of 
Figure~\ref{fig:cont50ev}.

A search with higher sensitivity could be performed exposing a dedicated
detector, whose feasibility study has not yet been worked out, at the future
neutrino beams from muon storage rings\cite{nufactory}. Such beams could
provide $\mathcal{O}(10^6)\nu_\mu$ charged-current events/year in a $10$ kg
fiducial mass detector, $1$ km away from the neutrino source. With this
statistic the sensitivity reached by Scenarios $C$ and $D$ could be exploited.

It is worthwhile observing that a high sensitivity search 
for $Z^{\prime}$, produced e.g.~via the processes 
$ gg \rightarrow q\bar{q} \rightarrow Z^{\prime}$, will be performed at 
LHC experiments (see for instance \cite{LHC}) few  years before neutrino 
factories will be operational.
Nevertheless, a negative result of such an analysis
would not decrease the interest of a high sensitivity search 
for $c \bar{c}$ production in neutrino interactions. 
An exotic $Z^{\prime}$ with stronger coupling to the $I_3 = 1/2$ 
component of weak isospin doublets could still give measurable effects at 
neutrino factories, unlike LHC experiments which are only sensitive 
to the $Z^{\prime}$ coupling to  charged leptons ($I_3 = -1/2$). 

\section{Conclusions}
We have presented a search for an extra neutral boson, $Z^{\prime}$,
by studying the associated charm production in neutral-current neutrino
interactions. The peculiarity of this process is that it involves only
second  generation fermions. Therefore, it allows the testing of the SM family 
universality through  the measurement of $Z^{\prime}$ couplings with the 
second family, a sector where so far there are no experimental limits.
We have also shown that, with  existing data, for the first time it is 
possible  to constrain the $Z^{\prime}$ couplings to the second 
generation.

\section*{Acknowledgements}
We are indebted to W.K.~Tung for useful discussions concerning 
the sea charm-parton component in the nuclei. We thank M.~Mangano for 
the helpful discussion.
We would also like to thank S.~Anthony for the careful reading of the 
manuscript.

\end{document}